\documentclass[review,3p]{elsarticle}
\usepackage{lineno,hyperref}
\modulolinenumbers[5]

\journal{EPSL}
\usepackage{lineno}
\biboptions{authoryear}

\usepackage{amsmath}
\usepackage{amssymb}
\usepackage{latexsym,bm}
\begin{document}

\begin{frontmatter}

\title{Ohmic dissipation in the Earth's outer core resulting from the free inner core nutation}


\author[mymainaddress,mysecondaryaddress]{Yufeng Lin \corref{mycorrespondingauthor}}
\ead{linyf@sustech.edu.cn}

\author[mysecondaryaddress]{Gordon Ogilvie}
\cortext[mycorrespondingauthor]{Corresponding author}

\address[mymainaddress]{Department of Earth and Space Sciences, Southern University of Science and Technology, Shenzhen 518055, China}
\address[mysecondaryaddress]{Department of Applied Mathematics and Theoretical Physics, University of Cambridge, Centre for Mathematical Sciences, Wilberforce Road, Cambridge CB3 0WA, UK}

\begin{abstract}
The diurnal tidal forces can excite a normal mode of the Earth's core, the free inner core nutation (FICN), which is characterized by a tilt of the rotation axis of the inner core with respect to the rotation axis of the outer core. The differential rotation between the  inner core and the outer core induces fluid motions in the outer core and gives rise to  Ohmic dissipation in the presence of the Earth's internal magnetic field. Nutation measurements can reflect such dissipation if it is sufficiently strong and thus can provide insights into the properties and dynamics of the Earth's core. 
In this study we perform a set of numerical calculations of the linear perturbations in the outer core induced by the FICN at very low Ekman numbers (as small as $10^{-11}$). Our numerical results show that the back-reaction of the magnetic field notably alters the structure and length scale of the perturbations induced by the FICN, and thus influences the Ohmic dissipation resulting from the perturbations. When the Ekman number is sufficiently small, Ohmic dissipation tends to be insensitive to the fluid viscosity and to the magnetic diffusivity, which allows us to estimate the Ohmic dissipation associated with the FICN without relying on an extrapolation. In contrast to the results of Buffett (2010b), the estimated Ohmic dissipation based on our numerical calculations is too weak to account for the observed damping of the FICN mode. This also implies that nutation measurements cannot provide effective constraints on the strength of the magnetic field inside the Earth's outer core.       
\end{abstract}


\end{frontmatter}
\section{Introduction}
The tidal forces exerted on the equatorial bulge by the Moon and the Sun cause variations in the orientation of the Earth's rotation axis, namely precession and nutation. Precession is a continuous change of the rotation axis around the ecliptic normal, while nutation consists of small periodic variations on top of precessional motion \citep{Mathews1992}.   

The frequencies and amplitudes of the nutations depend on internal structures and properties of the Earth. Therefore, observations of the Earth's nutations can provide independent constraints on the Earth's interior from other observations. Studies in this line can be dated back to \cite{Hopkins1839}, who investigated the fluidity of the interior of the Earth based on the observations of precession and nutation. Modern techniques such as very long baseline interferometry have provided high precision  measurements of the Earth's nutations for decades \citep{Herring1986,Mathews2002}, which can be used to constrain properties and dynamics of the Earth's interior, and couplings between different layers \citep{Buffett1992,Jackson2009,Buffett2010,Buffett2010a,Koot2010a,Koot2011}. This study mainly concerns Ohmic dissipation arising from nutation-induced fluid motions in the outer core, which may provide an extra damping mechanism of the Earth's nutations and can provide insights into the magnetic field within the Earth's core \citep{Buffett2010a}. 

The solid inner core (SIC) and the fluid outer core (FOC) are not locked to the nutations of the mantle owing to the existence of two diurnal normal modes, the free core nutation (FCN) and the free inner core nutation (FICN). The two modes are characterized by a tilt of the rotation axes of the FOC and the SIC with respect to the rotation axis of the mantle \citep{DeVries1991,Mathews1992}. The excitation of the FCN and the FICN modes by the diurnal tidal forces slightly enhances the amplitudes of the Earth's observed nutations; meanwhile, dissipative processes associated with these modes are evident in high precision measurements of nutation series as a phase lag between the tidal force and the nutational responses \citep{Mathews2002, Koot2010a}. The FICN is of particular interest to study the properties and dynamics of  the Earth's core as this mode mainly involves a relative rotation between the SIC and the FOC \citep{DeVries1991}. The damping of the FICN mode is usually attributed to the magnetic and viscous couplings at the inner core boundary (ICB) \citep{Mathews2002,Dumberry2012}. However, a strong radial magnetic field at the ICB or a large fluid viscosity of the outer core is invoked in order to explain the observed damping \citep{Buffett2002,Mathews2005,Koot2010a}. 

Apart from dissipation localized near the ICB, more distributed dissipative processes within the Earth's core may play a role in the damping of the FICN. \cite{Koot2011}  proposed that the viscoelastic deformation of the SIC,  together with the magnetic coupling at the ICB, can explain the observed damping of the FICN. Alternatively, \cite{Buffett2010a} demonstrated that the tilt of the  spheroidal inner core can induce internal shear layers in the FOC, which generate electrical currents by twisting the internal magnetic field, leading to an extra Ohmic dissipation in the outer core. Based on an extrapolation of a scaling from numerical results at moderate Ekman numbers, he argued that Ohmic dissipation in the FOC can explain the observed damping of the FICN, given a core-averaged magnetic field of 2.5 mT. This is also regarded as an indirect measurement of the strength of the magnetic field inside the outer core, which is crucial for our understanding of the geodynamo yet remains poorly constrained \citep{Gillet2010}. However, the scaling law used in \cite{Buffett2010a} did not take into account the back reaction of the magnetic field on the nutation-induced fluid motions (though a local correction due to the Lorentz force was applied after the extrapolation), which may result in an overestimate of Ohmic dissipation associated with the FICN as we shall show in this study.

Since the natural frequency of the FICN lies within the frequency range of inertial waves in a rotating fluid, the excitation of the FICN in turn generates inertial waves propagating in the fluid outer core, leading to conical shear layers around the characteristic surfaces of inertial waves  \citep{Hollerbach1995,Buffett2010a}. 
However, the presence of a magnetic field would alter the propagation of inertial waves and therefore their typical length-scale and associated dissipation, provided that the magnetic field is sufficiently strong or the Ekman number is sufficiently small \citep{Lin2018,Wei2018}. Given the extremely small Ekman number of the Earth's core, we expect non-negligible feedbacks of the magnetic fields on the nutation-induced inertial waves in the Earth's outer core.

In this study, we calculate the Ohmic dissipation associated with the FICN based on the model of \cite{Buffett2010a}, but taking into account the back-reaction of the magnetic field on the nutation-induced inertial waves. We perform a set of high resolution numerical calculations at extremely low Ekman numbers (as low as $10^{-11}$) and approach a regime in which Ohmic dissipation becomes insensitive to diffusive parameters. Our numerical results suggest that Ohmic dissipation within the outer core is too weak to account for the observed signatures of dissipation in the Earth's nutations. 

\section{The simplified model}
Our calculations are essentially based on the model used by \cite{Buffett2010a}, except that we solve the momentum equation and the magnetic induction equation simultaneously  in order to fully account the back-reaction of the magnetic field on the nutation-induced fluid motions. 

Both the FCN and the FICN modes should be excited by the diurnal tidal forces, leading to mutually misaligned rotation axes of the mantle $\bm{\Omega_m}=\Omega_0\bm{\hat{z}}$, the fluid outer core $\bm{\Omega_f}$ and the solid inner core $\bm{\Omega_s}$. Since our focus here is the FICN and the Ohmic dissipation associated with the FICN, we do not consider the differential rotation between the mantle and the outer core for simplicity.   
For small perturbations, the FICN can be regarded as a relative rotation of the solid inner core about an axis in the equatorial plane with respect to the outer core \citep{DeVries1991}. The relative rotation of the spheroidal inner core disturbs the fluid outer core and generates fluid motions in the bulk of the outer core. This effect can be approximated as a radial motion at a spherical inner core boundary in the frame that rotates with the outer core  \citep{Buffett2010a},

\begin{equation}\label{eq:ur}
u_r=-2\mathrm{i} \xi_s \varepsilon_s R_s \cos \theta \sin \theta \,\mathrm{e}^{\mathrm{i}(\phi-\omega t)},
\end{equation}  

where spherical coordinates ($r$, $\theta$, $\phi$) have been used with $\theta=0$ along the rotation axis of the outer core. Here $\xi_s$ is the amplitude of the differential rotation, $\varepsilon_s$ is the flattening of the inner core boundary, $R_s$ is the mean radius of the inner core and $\omega=-\Omega_o(1-\varepsilon_s)$ is the frequency of the FICN mode. The flattening is defined by $\varepsilon_s=(a-c)/a$, where $a$ and $c$ are the equatorial and polar radii of the inner core. We set $\varepsilon_s=0.0025$, which corresponds to the hydrostatic flattening of the inner core. The approximation of Eq. (\ref{eq:ur}) is valid to leading order in $\varepsilon_s$ and allows us to treat the outer core as a spherical shell  \citep{Buffett2010a}.

We consider the linear responses of the electrically conducting fluid shell to the radial motion of Eq. (\ref{eq:ur}) at the inner boundary in the presence of an ambient magnetic field. The structure of the magnetic field within the core is unknown. We simply assume a uniform vertical magnetic field $\bm{B_0}=B_0 \bm{\hat{z}}$ in this study, though other configurations of the ambient field will be briefly discussed in section \ref{sec:results}.

Using the radius of the Earth's core $R_c$, $\Omega_0^{-1}$, $B_0$ as units of length, time and magnetic field strength, the velocity and magnetic field perturbations $\bm u$ and $\bm b$ induced by the FICN are governed by the following linearized dimensionless equations in the frame rotating with the outer core \citep{Lin2018}:

\begin{equation} \label{eq:u}
-\mathrm{i}\omega \bm{u}+2\bm{\hat{z}}\times \bm u=-\nabla p+Le^2 (\nabla\times \bm b)\times \bm{B_0}+E_k \nabla^2 \bm u,
\end{equation}
\begin{equation}\label{eq:b}
 -\mathrm{i}\omega \bm{b}=\nabla\times( \bm u \times \bm{B_0})+E_m \nabla^2 \bm b,
\end{equation}
\begin{equation}\label{eq:div}
\nabla\cdot \bm u=0, \quad \nabla \cdot \bm b=0.
\end{equation}
The frequency $\omega=-(1-\varepsilon_s)$ is now non-dimensional. The dimensionless parameters in the above equations are the Lehnert number $Le$, the Ekman number $E_k$ and the magnetic Ekman number $E_m$ defined as

\begin{equation}
Le=\frac{B_0}{\sqrt{\rho \mu}\Omega_0 R_c}, \quad E_k=\frac{\nu}{\Omega_0 R_c^2}, \quad E_m=\frac{\eta}{\Omega_0 R_c^2},
\end{equation}
where $\rho$, $\mu$, $\nu$ and $\eta$ are density, magnetic  permeability,   fluid viscosity and magnetic diffusivity respectively. The ratio between $\nu$ and $\eta$ is known as the magnetic Prandtl number $Pm=\nu/\eta$. 

The Lehnert number is a dimensionless measurement of the strength of the ambient magnetic field. A magnetic field of 3mT corresponds to $Le\approx 10^{-4}$ for the parameters of the Earth's core. Following \cite{Buffett2010a}, we set $E_\eta=10E_k^{2/3}$, which gives rise to appropriate values of both $E_k\approx 10^{-15}$ and $E_\eta\approx 10^{-9}$ for the Earth's core as we decrease the Ekman number.  

As our focus is the dissipation in the bulk of the outer core, we adopt stress-free boundary conditions for the velocity perturbations and insulating boundary conditions for the magnetic perturbations at both the ICB and CMB to minimize dissipation at the boundaries. 

The perturbations lead to both viscous and Ohmic dissipation in the outer core. At small Ekman numbers, viscous dissipation is negligible in comparison with Ohmic dissipation. The dimensionless Ohmic dissipation rate is given as 

\begin{equation}
D_\eta=\frac{1}{2}Le^2E_m\int_V|\bm{\nabla\times b}|^2 \mathrm{d}V,
\end{equation}
where the integral is evaluated over the volume of the fluid domain.

\section{Numerical method}
The simplified model allows us to mimic the fluid motions induced by the FICN through a radially forced problem in a spherical shell. Equations (\ref{eq:u} - \ref{eq:div}), subject to the corresponding boundary conditions, are  solved numerically using a pseudo-spectral method based on a spherical harmonic expansion in a spherical shell. Both the velocity and magnetic field perturbations are expanded as 

\begin{equation} \label{eq:u_RST}
\bm u=	\sum_{l=1}^{L} u_l(r) \bm R_l+\sum_{l=1}^{L} v_l(r) \bm S_l+ \sum_{l=1}^{L} w_l(r) \bm T_l ,
\end{equation}
\begin{equation} \label{eq:b_RST}
\bm b=	\sum_{l=1}^{L} a_l(r) \bm {R}_l+\sum_{l=1}^{L} b_l(r) \bm S_l+ \sum_{l=1}^{L} c_l(r) \bm T_l,
\end{equation}
where

\begin{equation}
\bm{R}_l=Y_l^1(\theta,\phi) \bm{\hat{r}}, \ \bm{S}_l=r \bm{\nabla} Y_l^1(\theta,\phi) , \  \bm{T}_l=r \bm{\nabla \times}  \bm R_l.
\end{equation}
Here $Y_l^1 (\theta,\phi)$ is a spherical harmonic of degree $l$ and order $m=1$. As the background state is axisymmetric, the linear perturbations are decoupled for different $m$. For the FICN problem, only $m=1$ modes are involved as we can see from the boundary condition in equation (\ref{eq:ur}). Radial dependences are discretized using Chebyshev collocation on $N+1$ Gauss-Lobatto nodes. 

The radial motion in equation (\ref{eq:ur}) can be imposed as the following dimensionless boundary condition at the inner boundary:

\begin{equation}
u_l=4\mathrm{i}\xi_s\varepsilon_s\gamma\sqrt{\frac{2\pi}{15}}\delta_{l2},
\end{equation}
where $\xi_s$ is now the dimensionless amplitude of the differential rotation, and $\gamma=R_s/R_c=0.35$ is the relative size of the inner core. As we consider linear responses in this study, we set $\xi_s=1$ for simplicity but the results can be easily rescaled to arbitrary amplitudes. 

Note that the stress-free condition at the inner boundary needs to take into account the tangential stress due to the angular variation of the radial velocity at the boundary and is given as \citep{Ogilvie2009}

\begin{equation}
\frac{u_l}{r^2}+\frac{\mathrm{d}}{\mathrm{d}r}\left(\frac{v_l}{r}\right)=\frac{\mathrm{d}}{\mathrm{d}r}\left(\frac{w_l}{r}\right)=0.
\end{equation}
 
The non-penetration and stress-free conditions at the outer boundary lead to   

  \begin{equation}
u_l=\frac{\mathrm{d}}{\mathrm{d}r}\left(\frac{v_l}{r}\right)=\frac{\mathrm{d}}{\mathrm{d}r}\left(\frac{w_l}{r}\right)=0.
\end{equation}

For the magnetic field perturbations, an insulating boundary condition is used at both the ICB and CMB, which requires vanishing toroidal component, i.e. $c_l=0$, and to match a potential field for the poloidal component.

Projecting the governing equations onto spherical harmonics and imposing the corresponding boundary conditions generate a large block-tridiagonal linear system, which is solved using a direct solver. Details of the numerical scheme are referred to our previous study \citep{Lin2018}, in which the numerical scheme has been validated and used to study tidal dissipation in the presence of a magnetic field in a more general context.

We use a typical truncation of $L=600$ and $N=300$ for most calculations, but higher resolutions up to $L=1200$ and $N=600$ are also used for a few calculations at low Ekman nubmers ($E_k<10^{-10}$) and small Lehnert numbers ($Le<10^{-4}$). The convergence of the numerical solutions is examined by increasing the resolutions and by checking the spectra of the kinetic and magnetic energy.

\section{Results}\label{sec:results}

\begin{figure}
\begin{center}
\includegraphics[width=0.8 \textwidth]{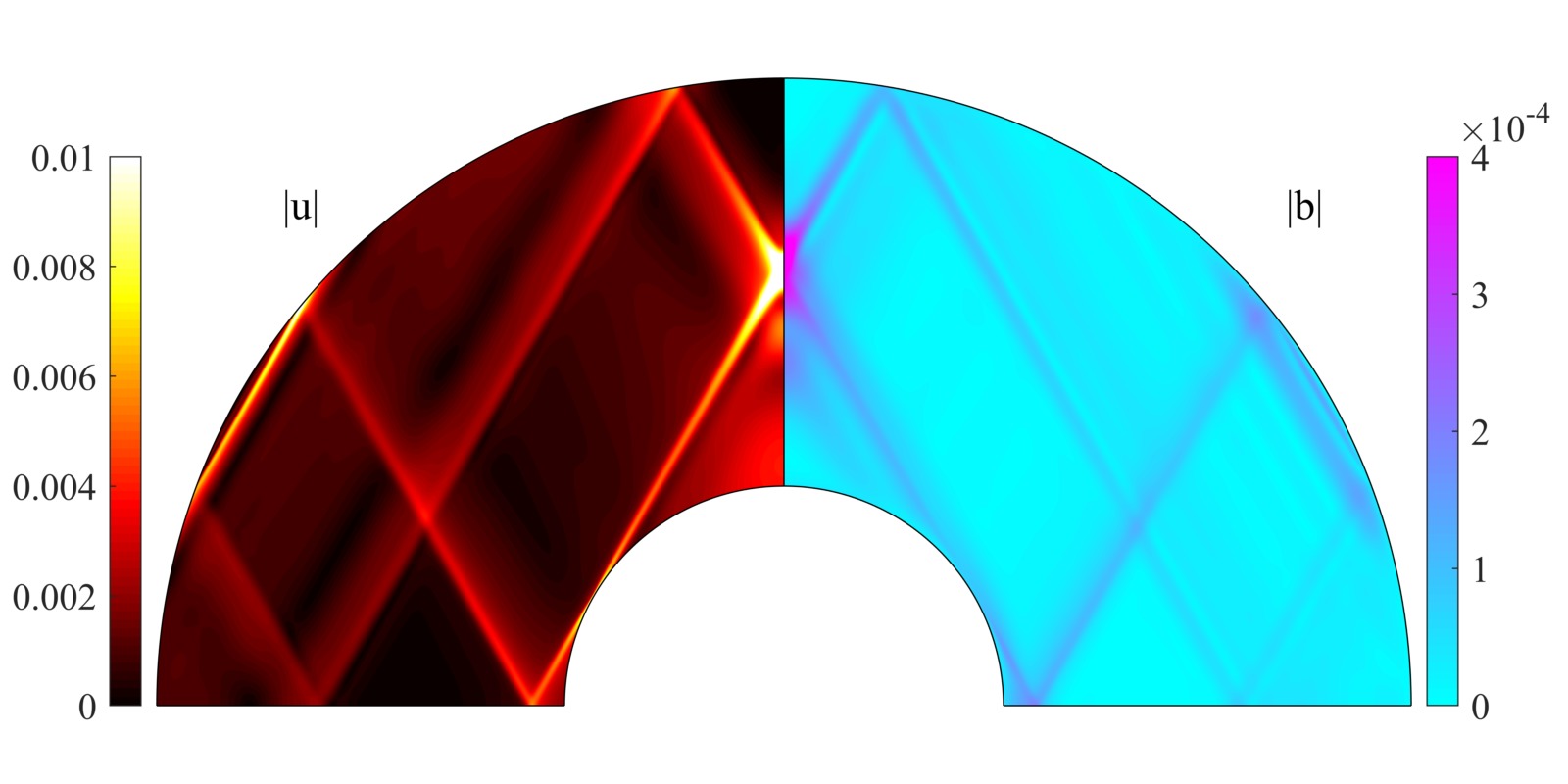}\\
\vspace*{-1cm}
(a)\\
\includegraphics[width=0.8 \textwidth]{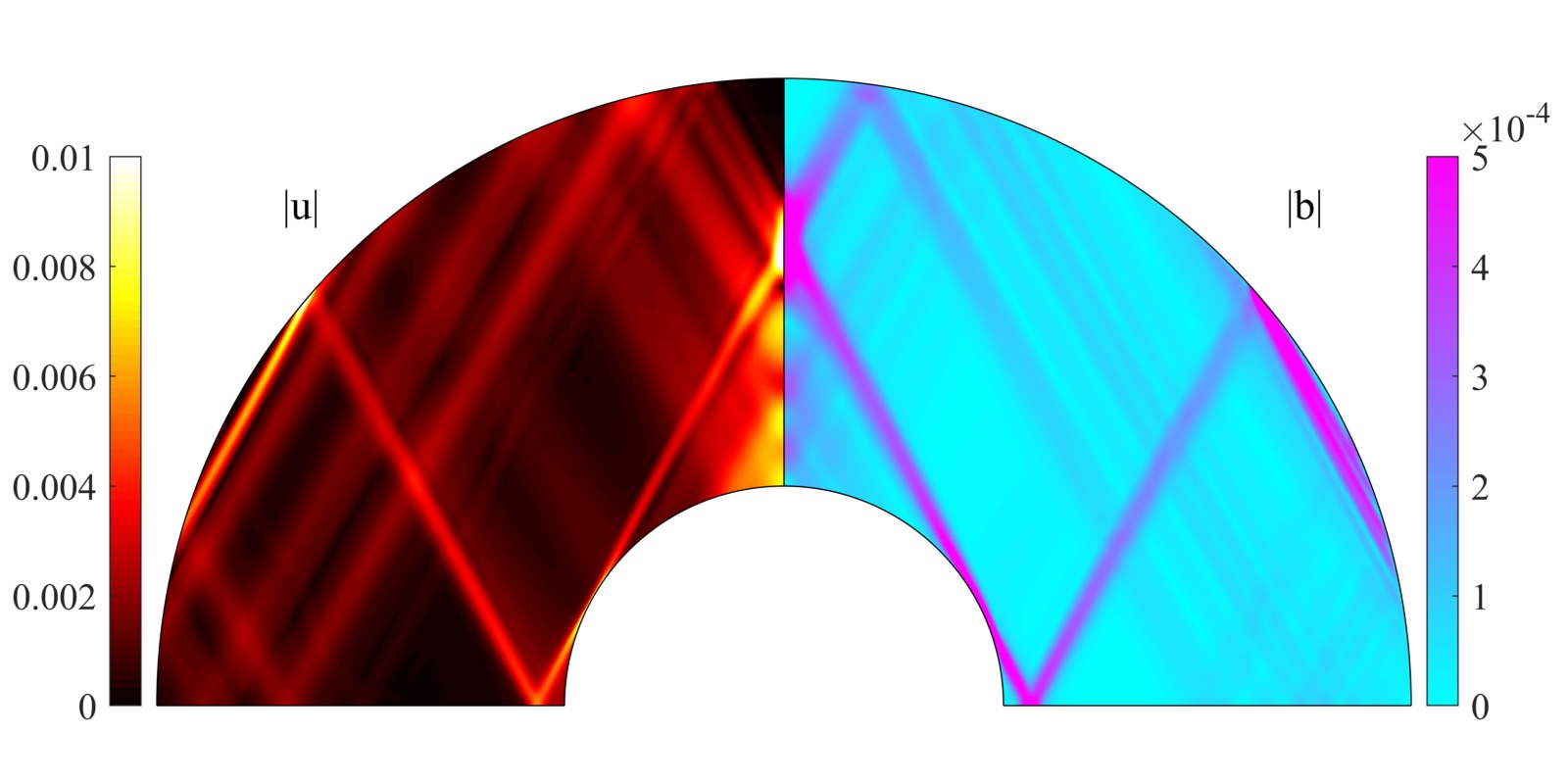}\\
\vspace*{-1cm}
(b)\\
\includegraphics[width=0.8 \textwidth]{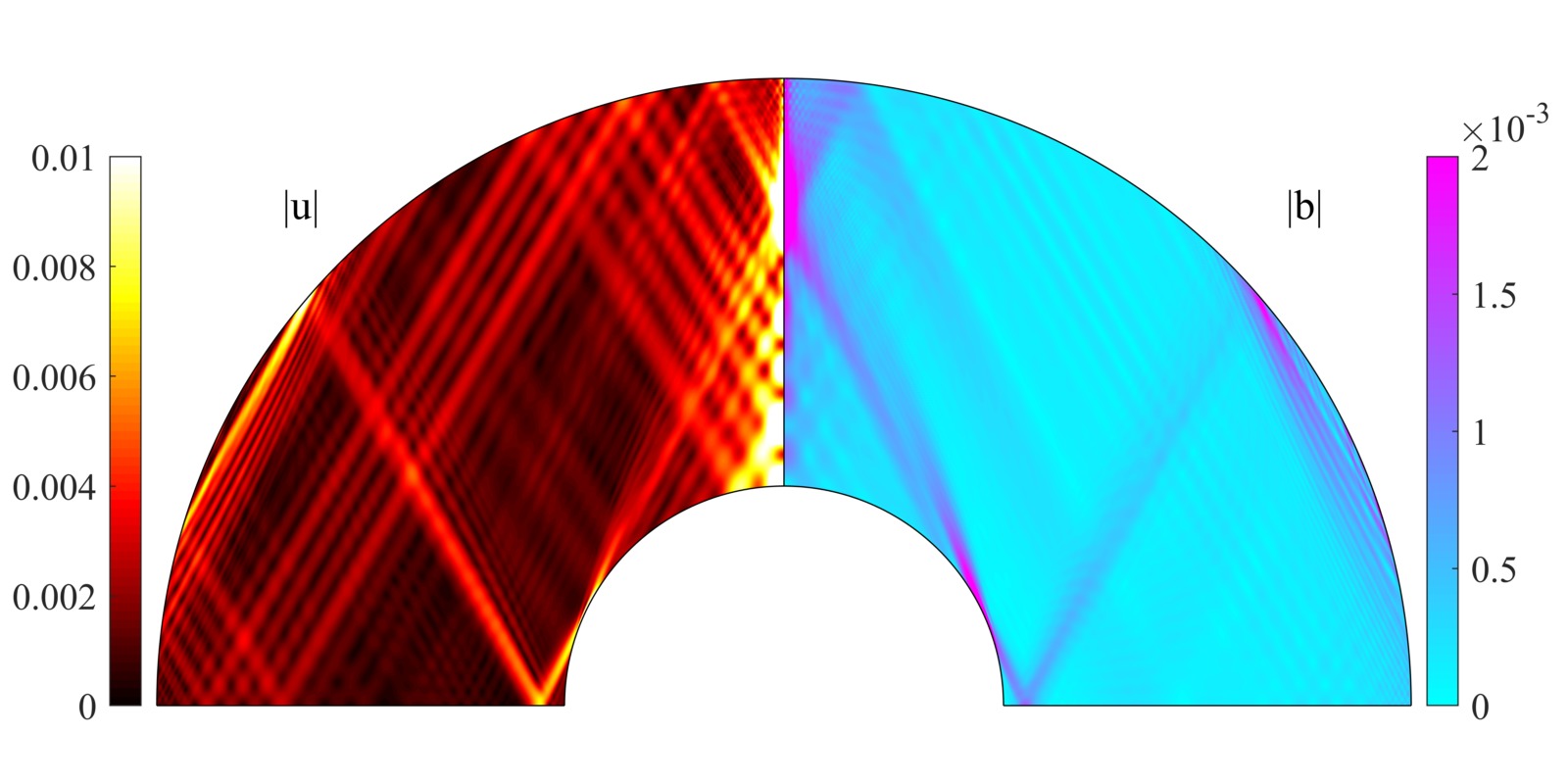} \\
\vspace*{-1cm}
(c)
\end{center}
\caption{Magnitude of the velocity $|\bm u|$ (left) and magnetic field $|\bm b|$ (right) in a meridional cross-section at different Ekman numbers with a fixed Lehnert number $Le=2.0\times 10^{-3}$. (a) $E_k=10^{-7}$; (b) $E_k=10^{-9}$; (c) $E_k=10^{-11}$. In each case, $E_\eta=10E_k^{2/3}$. }
\label{fig1}
\end{figure}

Fig. \ref{fig1} shows structures of the velocity and magnetic perturbations in a meridional cross-section at different Ekman numbers  
given a fixed Lehnert number $Le=2.0 \times 10^{-3}$. At $E_k=10^{-7}$ in Fig. \ref{fig1} (a), both velocity and magnetic field perturbations are concentrated in conical shear layers, which are known as inertial waves emanating from the critical latitude at the inner boundary of a spherical shell \citep{Kerswell1995,Hollerbach1995,Rieutord1997a}. The shear layers have a constant angle with respect to the rotation axis owing to the peculiar dispersion relation of inertial waves.  In this case, the magnetic field has negligible influence on the velocity perturbations, while the magnetic field perturbations are produced by the velocity perturbations in the conical shear layers through the magnetic induction equation (\ref{eq:b}). The perturbations in Fig. \ref{fig1}(a) are reminiscent of solutions obtained by neglecting the Lorentz force as in \cite{Buffett2010a}.  

As the Ekman number is decreased, the Lorentz force starts to alter the propagation of inertial waves. As we can see in Fig. \ref{fig1} (b-c), perturbations are scattered away from the concentrated shear layers and spread out in the whole domain at extremely low Ekman number $E_k=10^{-11}$. We note that the perturbations remain dominated by the Coriolis force like inertial waves, but the Lorentz force does play a part as some perturbations exhibit a slightly different angle with respect to the rotation axis. 
 
Our previous study  has shown that the magnetic field starts to influence the propagation of inertial waves and thus the corresponding dissipation when $Le>O(E_\eta^{2/3})$ in the limit of $P_m\ll1$ \citep{Lin2018}. As we set $E_\eta=10E_k^{2/3}$ (which guarantees $P_m\ll1$) in this study, the criterion can be represented in terms of Ekman number, i.e. $E_k\lesssim 10^{-3/2}Le^{9/4}$. 

\begin{figure}
\begin{center}
\includegraphics[width=0.6 \textwidth]{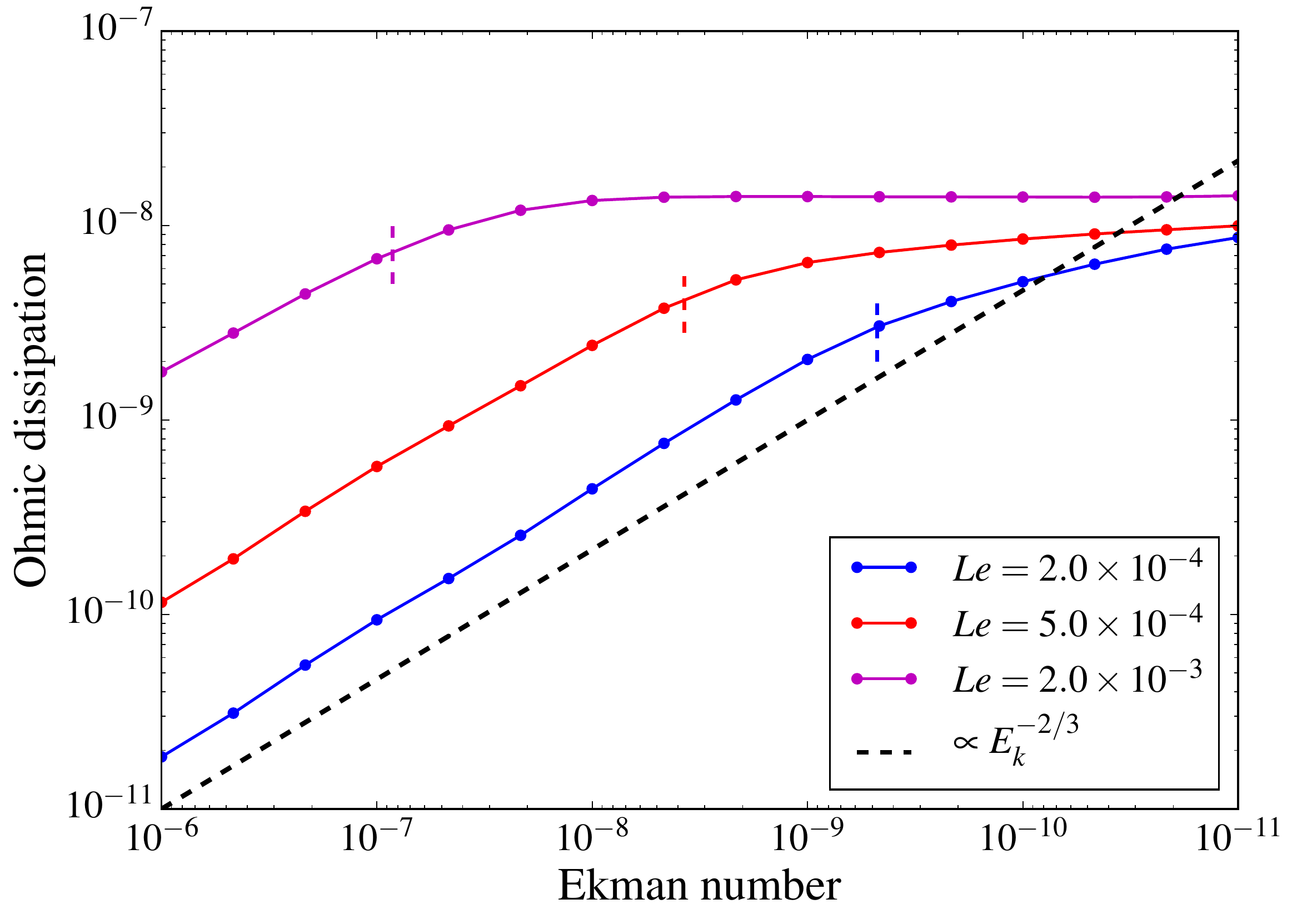}
\end{center}
\caption{Ohmic dissipation as a function of the Ekman number $E_k$ with different $Le$. Short vertical dashes lines indicate the critical Ekman number  $E_k= 10^{-3/2}Le^{9/4}$. In each case, $E_\eta=10E_k^{2/3}$. The Ekman number decreases from left to right.}
\label{fig:Ohmic_Ek}
\end{figure}

The relevance of this criterion is clearly evident in Fig. \ref{fig:Ohmic_Ek}, in which  the Ohmic dissipation is plotted as a function of the Ekman number at various Lehnert numbers. Short dashed lines in Fig. \ref{fig:Ohmic_Ek} indicate the critical $E_k$ for a given $Le$. When the Ekman number is larger than the critical value, Ohmic dissipation follows $D_\eta \propto Le^2E_k^{-2/3}$, which is the expected scaling when the Lorentz force is neglected \citep{Buffett2010a}. 
 However, this scaling  breaks down when the Ekman number is smaller than the critical value as we can see from Fig. \ref{fig:Ohmic_Ek}. More importantly, Ohmic dissipation tends to become independent of the Ekman number  when $E_k$ is sufficiently small. If the scaling $D_\eta\propto Le^2E_k^{-2/3}$ is used to extrapolate to the condition of the Earth's core, Ohmic dissipation  would be significantly over estimated. 
 \begin{figure}
\begin{center}
\includegraphics[width=0.49 \textwidth]{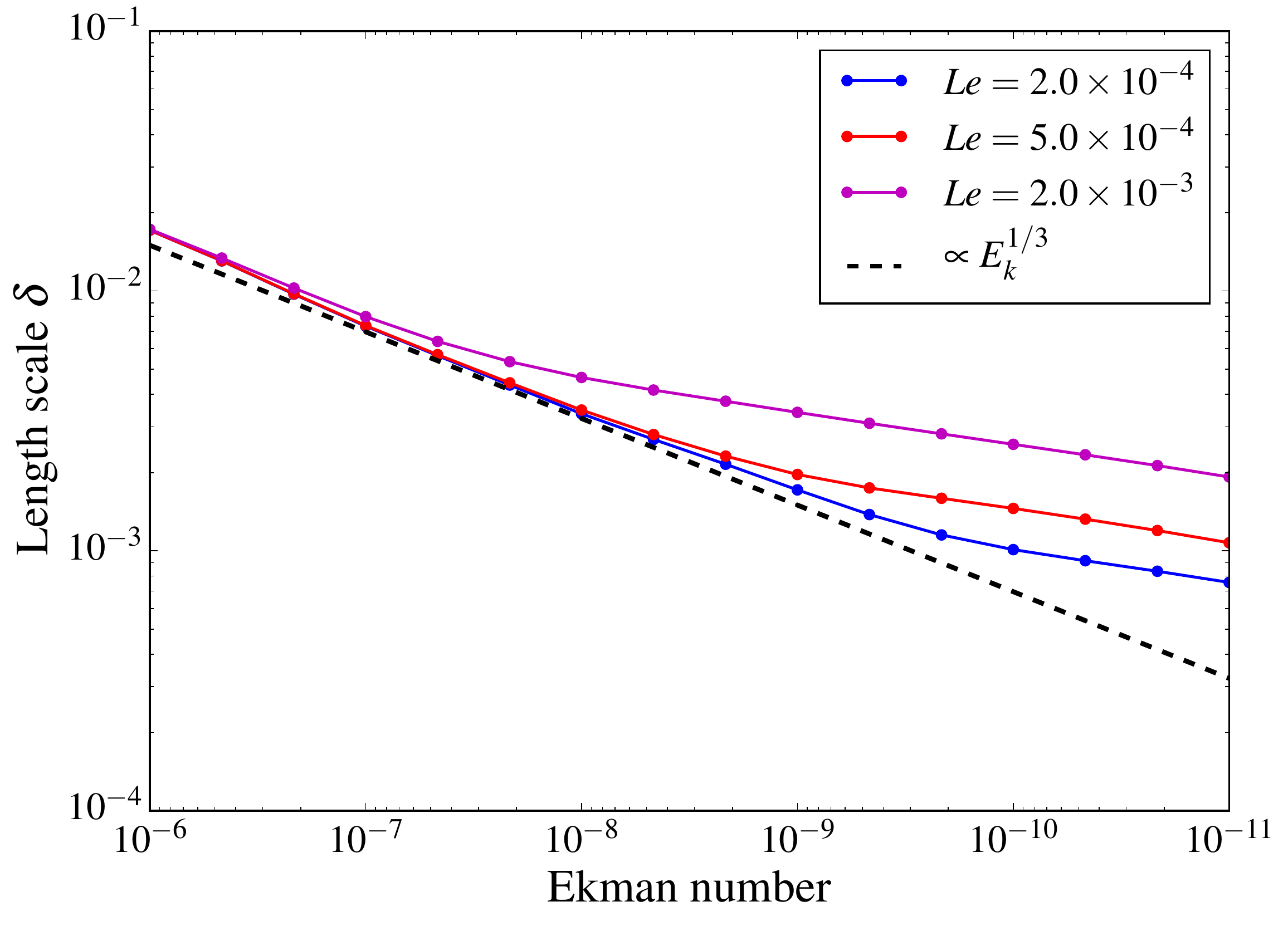}
\includegraphics[width=0.49 \textwidth]{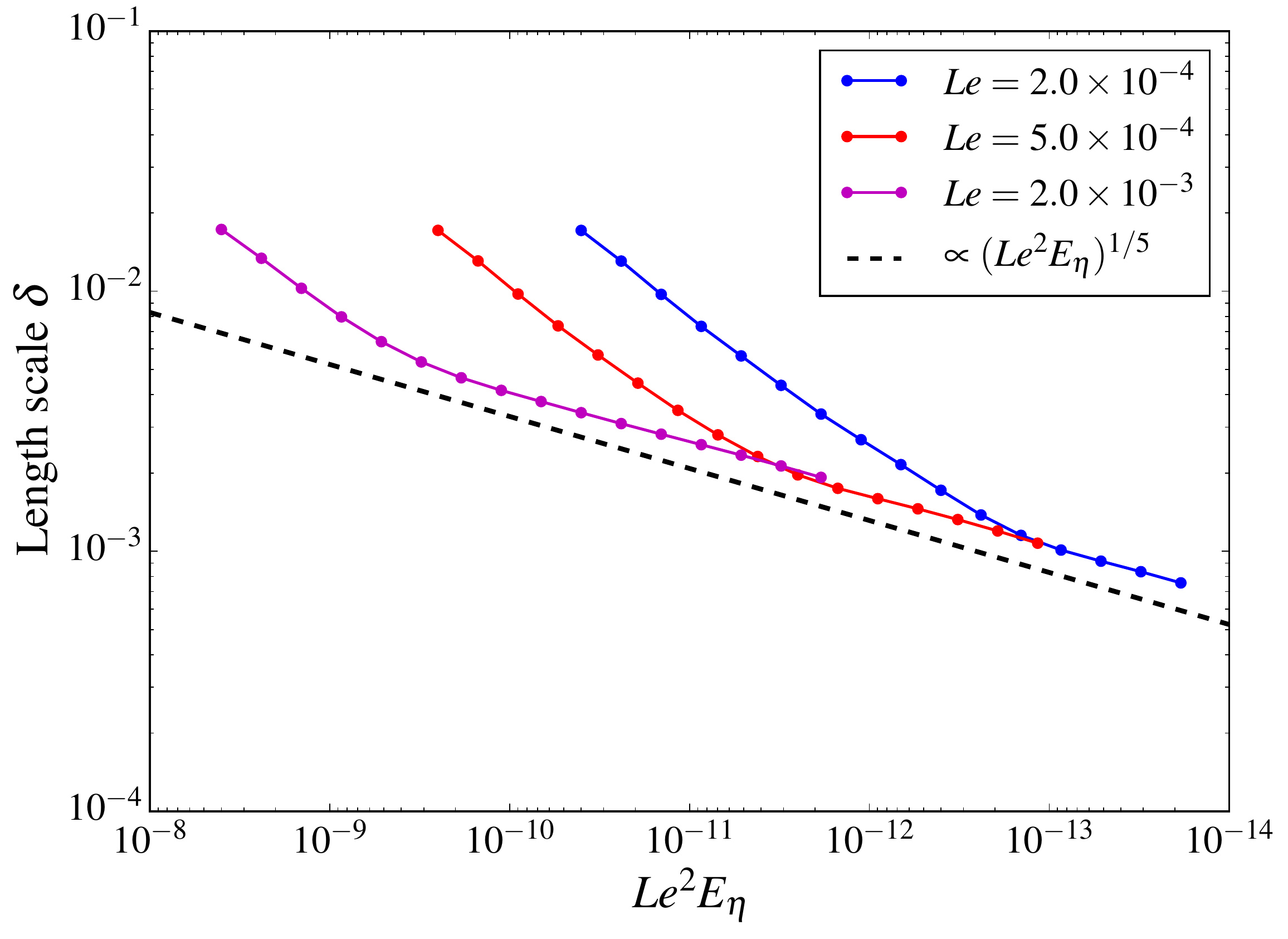}\\
(a) \hspace*{0.5 \textwidth} (b)
\end{center}
\caption{ Length scale $\delta$ as a function of (a) Ekman number $E_k$ and (b) $Le^2E_\eta$ with different $Le$.}
\label{fig_length}
\end{figure}

The Ohmic dissipation scaling $D_\eta\propto Le^2E_k^{-2/3}$ breaks down because the magnetic field modifies the length scale of the perturbations when $E_k<O(Le^{9/4})$. Fig. \ref{fig_length} shows the length scale of the magnetic  perturbations as a function of the Ekman number. The length scale $\delta$ is estimated by the ratio between the magnetic energy and the Ohmic dissipation:

\begin{equation}
\delta^2=\frac{\int_V|\bm{b}|^2 \mathrm{d}V}{\int_V|\bm{\nabla\times b}|^2 \mathrm{d}V}.
\end{equation}

If the Lorentz force is neglected, the length scale is expected to be of  $\delta=O(E_k^{1/3})$ \citep{Kerswell1995, Rieutord1997a, Buffett2010a}. However, we can see from Fig. \ref{fig_length}(a) that the length scale deviates from $\delta=O(E_k^{1/3})$ as the Ekman number is decreased. The length scale varies less steeply as $E_k$ and becomes dependent on the strength of the magnetic field ($Le$), suggesting a significant influence of the magnetic field on the length scale. Indeed, Fig. \ref{fig_length}(b) shows that the length scale tends to scale as $\delta\sim (Le^2E_\eta)^{1/5} $ when the Lorentz force becomes significant. We can see that $Le^2E_\eta$ is a relevant control parameter by substituting Eq. (\ref{eq:b}) into Eq. (\ref{eq:u}), but unfortunately we cannot provide a rigorous explanation for the exponent of $1/5$. Anyway, our numerical results show that the magnetic field broadens the length scale of the perturbations. 
In fact, the relatively larger length scale allows us to perform numerical calculations at very low Ekman numbers. Otherwise, it would be very demanding to numerically resolve the length scale of $O(E_k^{1/3})$ at $E_k=10^{-11}$.

In short, Figs. \ref{fig:Ohmic_Ek} and \ref{fig_length} clearly show that the scaling laws obtained when the Lorentz force is neglected are inappropriate to extrapolate to the parameter regime of the Earth's core. Such an extrapolation would lead to an overestimate of the dissipation in the outer core resulting from the FICN.

\begin{figure}
\begin{center}
\includegraphics[width=0.6 \textwidth]{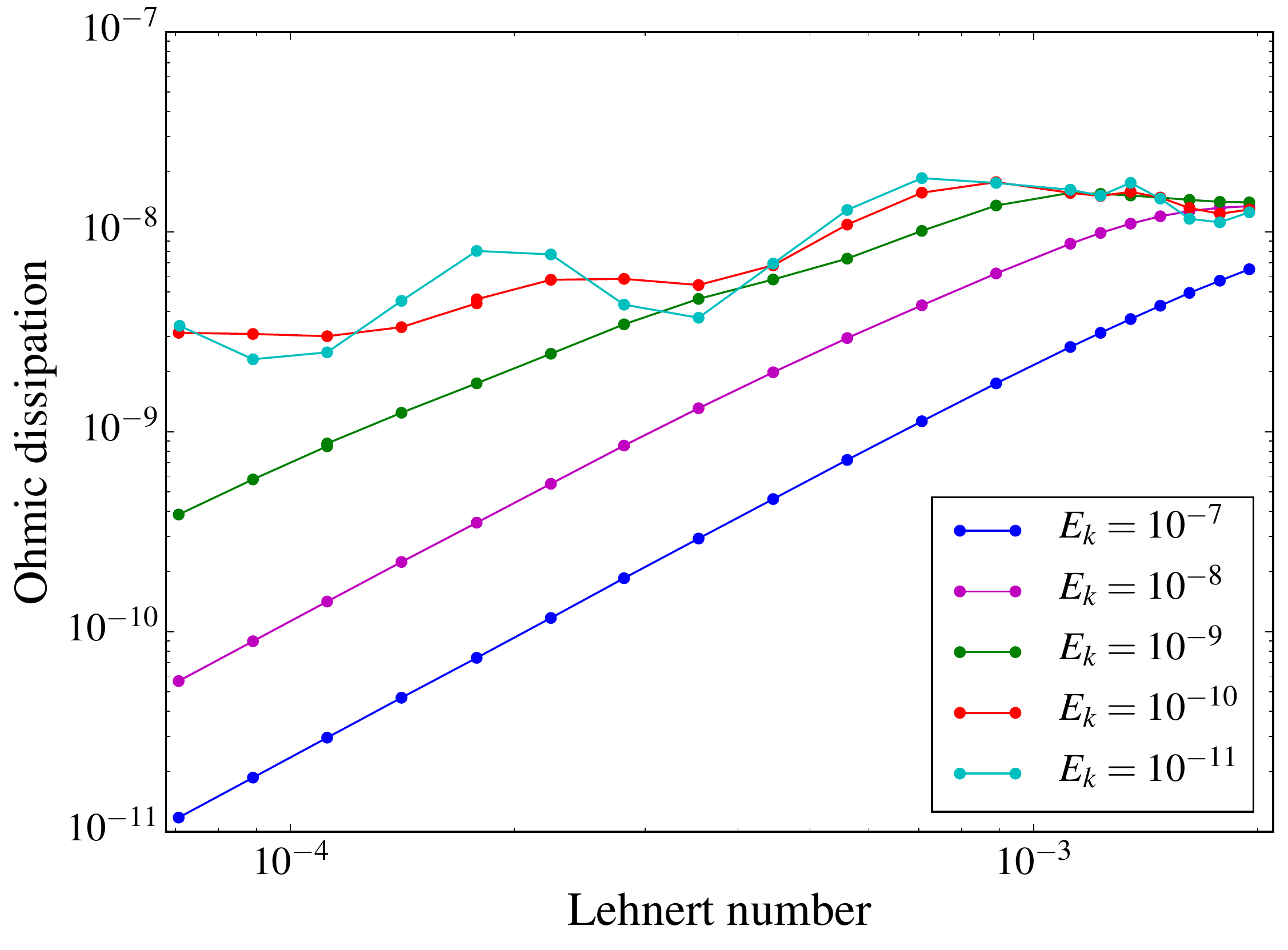}\\
\end{center}
\caption{Ohmic dissipation as a function of the Lehnert number at different Ekman numbers.}
\label{fig_Le}
\end{figure}
Fig. \ref{fig_Le} shows the Ohmic dissipation as a function of the Lehnert number at various Ekman numbers. At relatively large $E_k$ and small $Le$, Ohmic dissipation increases with increasing $Le$ or decreasing $E_k$, i.e. $D_\eta\propto Le^2E_k^{-2/3}$ as we have shown in Fig. \ref{fig:Ohmic_Ek}. When the magnetic field starts to influence the perturbations, Ohmic dissipation becomes insensitive to the diffusive parameters and does not vary monotonically as $Le$. The presence of a magnetic field modifies the dispersion relation of inertial waves, leading to magnetic Coriolis waves, whose frequency depends on the strength of the magnetic field. Troughs and peaks in Fig. \ref{fig_Le} may be related to weak resonances of the magnetic Coriolis waves with the tidal frequency \citep{Lin2018}.   
\begin{figure}
\begin{center}
\includegraphics[width=0.6 \textwidth]{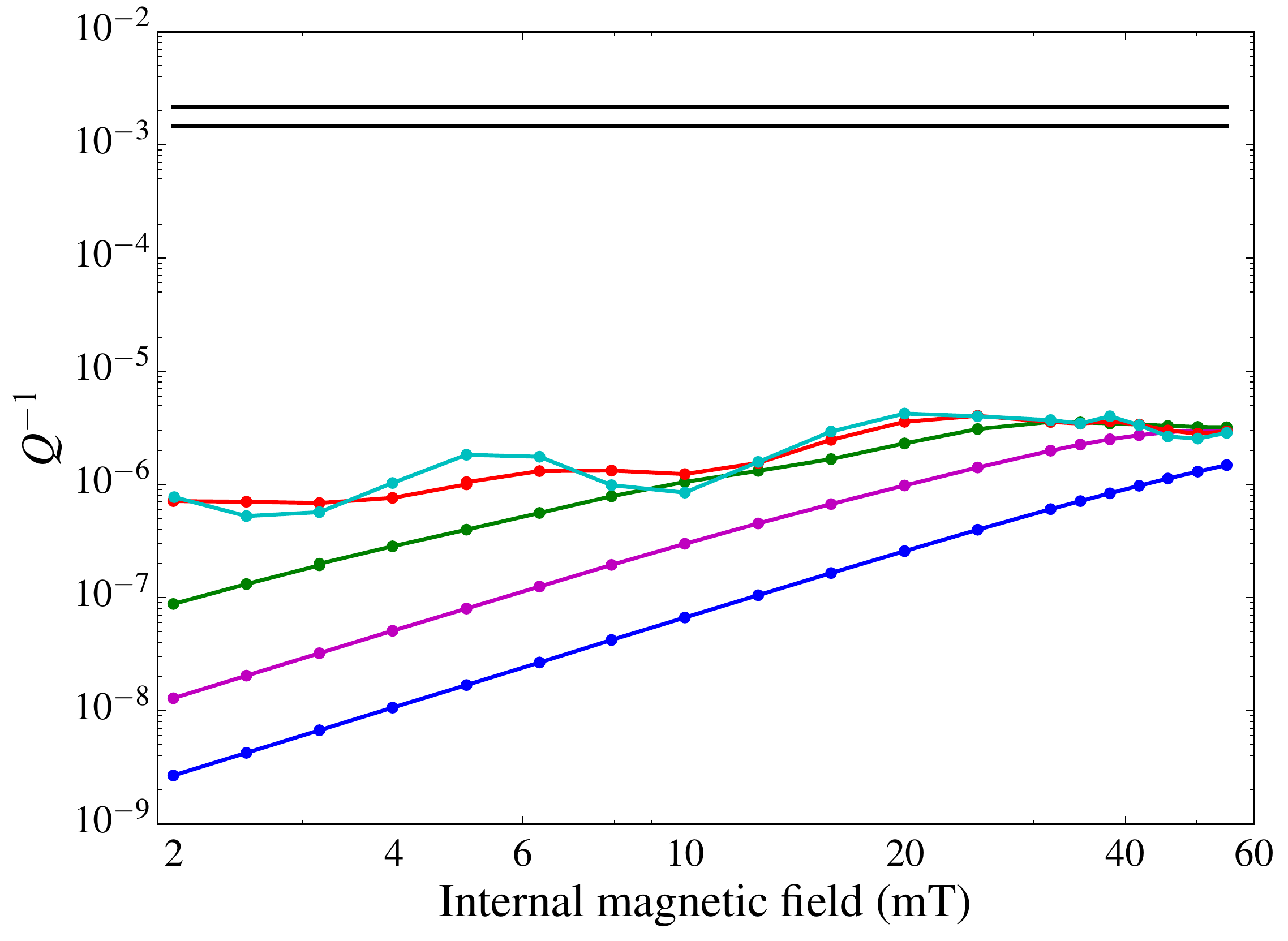}\\
\end{center}
\caption{Inverse of the quality factor $Q^{-1}$ as a function of the magnetic field strength at different Ekman numbers. Black lines represent the observed quality factor of the FICN mode, $Q=667$ \citep{Mathews2002} and $Q=459$ \citep{Koot2010a}. Other symbols are the same as in Fig. \ref{fig_Le}.}
\label{fig_Q}
\end{figure}

Finally, we compare the estimated Ohmic dissipation based on our calculations with the observed damping of the Earth's nutation. Energy dissipation resulting from tidal forces is often parameterized as the so-called quality factor $Q$, which is the ratio  between the energy stored in the tidal response and the energy dissipated per tidal cycle \citep{Goldreich1963}.  For the FICN, the energy stored in the tidal response is mainly the relative rotation of the SIC with respect to the FOC. Ohmic dissipation resulting from the FICN can be represented as the quality factor 

\begin{equation}
Q=\frac{\omega K_s}{D_\eta},
\end{equation}
where $K_s$ is the kinetic energy of the relative rotation of the SIC. Note that $Q$ is independent of the amplitude of the differential rotation $\xi_s$ for linear perturbations as both $K_s$ and $D_\eta$ are proportional to $\xi_s^2$.

The graph of Ohmic dissipation versus Lehnert number in Fig. \ref{fig_Le}  is converted into a graph of the inverse quality factor $Q^{-1}$ versus the magnetic field strength in Fig. \ref{fig_Q}. The observed dissipation can be also related to the quality factor of the FICN \citep{Buffett2010a}, which is represented as horizontal black lines in Fig. \ref{fig_Q}. We can see that the quality factor $Q$ due to Ohmic dissipation is about three orders of magnitude larger than the observed $Q$ even with a very strong magnetic field of  up to 50mT, meaning that Ohmic dissipation resulting from the perturbations in the FOC is too small to account for the observed damping of the FICN mode.

It is computationally demanding to access both low $E_k$ and $Le$. The smallest Lehnert number ($Le=7\times 10^{-5}$) used in this study corresponds to a magnetic field of 2mT. Within the range of parameters we could reach, our calculations have  shown that the Ohmic dissipation becomes insensitive to the fluid viscosity  provided that $E_k$ is sufficiently small, i.e. $E_k\leq 10^{-10}$. Therefore, the calculated dissipation can be applied to the conditions of Earth's core without relying on an extrapolation. 
 
In the scenario of a weak magnetic field ($\ll$2mT), though this is probably not the case for the Earth's core according to other estimates of the field strength \citep{Gillet2010,Hori2015}, or a high viscosity (e.g. eddy viscosity), Ohmic dissipation remains too   small to match the observed damping, even using the scaling $D_\eta\propto Le^2E_k^{-2/3}$ \citep{Buffett2010a}. 

Fig. \ref{fig_Q} also implies that the observed damping of the FICN cannot provide effective constraints on the magnetic field strength within the Earth's outer core, at least not within our simplified model. However, nutation measurements may still be used to constrain the radial magnetic field at the CMB and ICB through electromagnetic couplings at the fluid-solid interfaces \citep{Buffett2002,Koot2010a}, which are not considered in this study. 

We also performed a few calculations using the no-slip boundary condition, which show that the no-slip condition has little influence on Ohmic dissipation but increases the viscous dissipation due to the thin viscous boundary layer. However, viscous dissipation decreases on reducing the Ekman number and becomes negligible in comparison with Ohmic dissipation at smaller Ekman numbers.

Regarding the structure of the ambient magnetic field, we have also considered an axial dipolar magnetic field, but the results are qualitatively similar to that  of a uniform vertical field, so we do not present detailed results here. We also considered a simple toroidal field $\bm{B_0}=B_0r\sin\theta \bm{\hat \phi}$, which is proposed by \cite{Malkus1967}. For Malkus' toroidal field, Ohmic dissipation is much smaller than that of a poloidal field with similar strength. We should note that Malkus' field is special because the magnetohydrodynamic problem can be reformulated as a modified inertial waves problem \citep{Malkus1967}. Furthermore, \cite{Buffett2010a} has pointed out that nutation-induced fluid motions are not sensitive to the azimuthal component of the magnetic field. However, these considerations are all about large scale axisymmetric magnetic fields. The effects of non-axisymmetric and short-wavelength magnetic fields would be worth studying in great detail \citep{Buffett2007,Dumberry2012}.  

\section{Conclusion} \label{sec:conclusion}
Using a simplified model based on \cite{Buffett2010a}, we calculated the linear perturbations induced by the FICN mode in the outer core, by taking into account the back reaction of the internal magnetic field on the fluid motions. Ohmic dissipation resulting from these perturbations was estimated and compared with the observed damping of the FICN. 

Our calculations at very low Ekman numbers (as small as $10^{-11}$) have shown that magnetic fields alter the structure and the length scale of the perturbations. Neglecting the Lorentz force in the momentum equation can lead to a significant overestimate of Ohmic dissipation resulting from the FICN. Our numerical results suggest that Ohmic dissipation becomes insensitive to the fluid viscosity (and magnetic diffusivity) at low Ekman numbers, which allows us to estimate Ohmic dissipation associated with the FICN without relying on an extrapolation. The estimated Ohmic dissipation based on our calculations is too small to account for the observed damping of the FICN mode. Viscous and magnetic couplings at the inner core boundary and viscoelatic deformation of the inner core may be responsible for the observed damping, but improved models on these mechanisms are required to better explain the nutation measurements \citep{Koot2010a,Dumberry2012}.  

As the damping of the FICN due to Ohmic dissipation in the outer core is too weak to be detectable, nutation measurements cannot provide effective constraints on the magnetic field strength within the Earth's outer core based on the model used here. The estimate of the core-averaged magnetic field obtained by \cite{Buffett2010a} using the nutation measurements is affected by an overestimate of Ohmic dissipation as we have shown. However, we note that Earth's nutations, through the magnetic coupling at boundaries, may still provide useful insights into magnetic fields at the ICB and CMB, which are inevitably related to the  magnetic field in the bulk of fluid outer core.

In this paper, we consider only linear perturbations induced by the FICN. However, these perturbations could become unstable through parametric instabilities \citep{Kerswell1994,Lin2015}, which give rise to extra viscous and Ohmic dissipation. Nonlinear effects on the damping of the Earth's nutations remain to be explored.   

\section*{Acknowledgements}
YL acknowledges the support of the Swiss National Science Foundation through an advanced Postdoc.Mobility fellowship in Cambridge. 
This study is also supported by Center for Computational Science and Engineering of Southern University of Science and Technology.
\section*{References}


\end{document}